\documentclass[prd,preprintnumbers,showpacs,12pt]{revtex4}
\usepackage[active]{srcltx}
\usepackage{epsfig}
\usepackage{graphicx}
\usepackage{bm}
\usepackage{amsmath}
\usepackage{amssymb}

\newcommand{\be}{\begin{equation}}

\newcommand{\ee}{\end{equation}}
\newcommand{\bea}{\begin{eqnarray}}
\newcommand{\eea}{\end{eqnarray}}

\newcommand{\nn}{\nonumber}
\newcommand{\de}{\partial}

 \def\slash#1{\setbox0=\hbox{$#1$}#1\hskip-\wd0\dimen0=5pt\advance
       \dimen0 by-\ht0\advance\dimen0 by\dp0\lower0.5\dimen0\hbox
         to\wd0{\hss\sl/\/\hss}}
\usepackage{color}

\begin{document}
\hfill{\bf ICCUB-14069~~~\,}\vskip0.1cm

\title{Conformal symmetry for relativistic point particles: an addendum}

\author{Roberto Casalbuoni}\email{casalbuoni@fi.infn.it}
\affiliation{Department of Physics and Astronomy, University of Florence and
INFN, Florence, Italy}
\author{Joaquim Gomis} \email{gomis@ecm.ub.es}
\affiliation{
 Departament d'Estructura i Constituents de la
Mat\`eria and Institut de Ci\`encies del
Cosmos, Universitat de Barcelona, Diagonal 647, 08028 Barcelona,
Spain}

\begin{abstract}

We extend the results of our previous work on the conformal invariant description of two relativistic point particles. We consider here the most general lagrangian by using a conformal tensor $h_{\mu\nu}$, transforming as a Wilson line, and that allows us to construct invariant expressions for velocities taken at two different space-time points.

\end{abstract}
\pacs{11.25.Hf, 11.30.-j, 11.10.Ef, 03.30.+p}

\maketitle

\section{Introduction}\label{sec:0}

In a previous paper \cite{Casalbuoni:2014ofa}, we have studied a conformal invariant model for two interacting relativistic particles. Our hypothesis was that the theory could be described in terms of the four-velocities of the two particles and the relative distance, assuming to form Lorentz invariant objects by contracting the vectors by using the Minkowski metric tensor. It turned out that the action was uniquely determined. In the present note we want to enlarge this framework by using a tensor $h_{\mu\nu}$, introduced by  \cite{Boulware:1970ty} (see also \cite{Ferrara:1973eg}). This tensor appears naturally in conformal field theory in the two-point function of two vector currents and transforms as a Wilson line under special conformal transformations. Therefore it is possible to construct conformal scalars by contracting the velocities of the two particles taken at different points. We will show that the more general action depends on one arbitrary function of a conformal and diffeomorphism invariant dimensionless variable.  The diffeomorphism (Diff) invariance is the invariance under repametrization of the single parameter describing the  trajectories of the two particles.

In Section II we review some properties of the conformal group with a particular focus on the role of the discrete inversion transformation. In fact, the special conformal transformations can be recovered by using translation and inversions. In particular, we will show the  transformation properties of the tensor $h_{\mu\nu}$ under inversion.

In Section III we write down the most general conformal and Diff invariant action for two particles, showing that it depends on an arbitrary function of a dimensionless variable depending on the tensor $h_{\mu\nu}$. Furthermore, we reformulate the model by introducing two einbein variables. It turns out that the interaction term depends on a function of two dimensionless variables, with one depending on the einbeins. By a general argument these two formulations are equivalent, that is eliminating the einbeins from the second formulation one is bound to recover the first one. However we have been able to show it explicitly, only in the case in which the the arbitrary function is a power in the einbeins. We show also the relation with the model presented in \cite{Casalbuoni:2014ofa}, and we introduce another very simple model using the tensor $h_{\mu\nu}$. Eventually, we evaluate the first class constraints, related to the Diff invariance, of the last two models. In particular we find that the constraint arising from the second model is quadratic in the momenta, whereas in the first one is quartic.

In Section IV we make a general study of the first class constraint associated to Diff invariance. 
In the general case the constraint cannot rewritten explicitly, differently from the cases discussed in Section III.
However,  we are able to show the procedure that should be followed for any choice of the arbitrary function appearing in the version of the action not depending on the einbeins. We discuss also the particular case in which this function reduces to an arbitrary power. Is then possible to bring down the problem to the solution of an algebraic equation of the third degree in the dimensionless variable appearing in the arbitrary function.

Section V is devoted to the outlook and conclusions.

\section{Some properties of the conformal transformations}\label{sec:1}

The action of the special conformal transformations on the space-time coordinates (with a space-time of dimension different from 2) is given by
   \be \bar x_\mu=\frac{(x^\mu+c^\mu x^2)}{1+2c\cdot x+c^2 x^2}=\frac{(x^\mu+c^\mu x^2)}{\sigma(x)},~~~~\sigma(x)=1+2c\cdot x+c^2 x^2\label{eq:1}\ee
   An infinitesimal transformation is given by
   \be
   \delta x^\mu= -2(c\cdot x) x^\mu+c^\mu x^2=x^2h^{\mu\nu}(x) c_\nu\label{eq:3}\ee
   where
   \be
   h^{\mu\nu}(x)=g^{\mu\nu}-2\frac{x^\mu x^\nu}{x^2},\label{eq:4}\ee
   (we use the mostly plus metric $g_{\mu\nu}=(-,+\cdots,+)$), with the property
   \be
   h^{\mu\nu}(x)h_{\nu\lambda}(x)=g^\mu_\lambda\label{eq:5}\ee
   We recall that the generic special conformal transformation can be obtained through the following combination of transformations: (inversion)$\otimes$(translation)$\otimes$(inversion), where the inversion is defined as
   \be
x^\mu\to {\bar x}^\mu =\frac{x^\mu}{x^2}\label{eq:6}\ee
It follows that, in order to implement conformal invariance, it is enough to require Poincar\'e and inversion invariance. 
    Let us start with the transformation properties of $d x^\mu$. We have
    \be
    dx^\mu\to d\bar x^\mu= \frac{\de \bar x^\nu}{\de x^\nu}dx^\nu =\frac 1{x^2}h^\mu_\nu(x) dx^\nu\label{eq:7}\ee
    From this
    \be
 d\bar x^2=\frac{dx^2}{x^4}\label{eq:8}\ee
  Then, let us  consider the transformation properties of the derivative
    \be
       \frac{\de}{\de  x^\mu}\to \frac{\de}{\de \bar x^\mu}=\frac{\de x^\nu}{\de\bar x^\mu}\frac{\de}{\de x^\nu}=x^2 h_\mu^\nu(x)\frac{\de}{\de x^\nu}\label{eq:9} \ee
Notice that $h_{\mu\nu}(x)$ enjoys the following property $h^{\mu\nu}(\bar x)=h^{\mu\nu}(x)\label{eq:10}$.
    
Since we are interested in a two-particle model, a quantity of interest is the relative distance $r^\mu=x_1^\mu-x_2^\mu$. Under inversion we have

    \be \bar r^\mu=\bar x_1^\mu-\bar x_2^\mu=\frac {x_1^\mu}{x_1^2}-\frac{x_2^\mu}{x_2^2}=
    \frac{x_1^\mu x_2^2-x_2^\mu x_1^2}{x_1^2 x_2^2}\label{eq:11}\ee
    From which
    \be
    \bar r^2= \frac{r^2}{x_1^2 x_2^2}\label{eq:12}\ee
   
More generally, since the special conformal transformations depend on the space-time point, in order to define the scalar product among vectors defined at different points, we need the notion of parallel transport.  This problem has been solved in 1970 in \cite{Boulware:1970ty}. Let us consider the square of the distance between two points, from eq. (\ref{eq:12}) we see  that under inversion $r^2$ has a simple scaling behavior. Then, let us consider its logarithm, which transforms as
    \be
    \log(x_1-x_2)^2 \to   \log(x_1-x_2)^2-\log( x_1^2) -\log(x_2^2)\label{eq:13}   \ee
    The additional terms do not contribute by taking the double derivative with respect to the coordinates of the two points. Furthermore, in order to get a dimensionless  tensor  we multiply the result by $r^2$. That is, we define the symmetric conformal tensor
    \be
    h_{\mu\nu}(r)=-\frac 12r^2\frac{\de}{\de x_1^\mu}\frac{\de}{\de x_2^\nu}   \log(x_1-x_2)^2= g_{\mu\nu}-2\frac{r_\mu r_\nu}{r^2}\label{eq:14}\ee 
    Notice that  $h^{\mu\nu}(r)$  is formally equal to the $h$-tensor introduced previously (for this reason we use the same symbol). It follows that also $h^{\mu\nu}(r)$ satisfies the identity 
    \be h_{\mu\nu}(r)h^{\nu\rho}(r)=g_\mu^\rho\label{eq:15}\ee
   
    Let us now evaluate the properties of $h_{\mu\nu}$ under  inversion. From its very definition we have
    \be
    h_{\mu\nu}(r)\to \bar h_{\mu\nu}(\bar x)=-\frac 12\bar r^2\frac{\de}{\de \bar x_1^\mu}\frac{\de}{\de \bar x_2^\nu}   \log\bar r^2= h_\mu^\lambda(x_1)h_\nu^\rho(x_2)h_{\lambda\rho}(r)\label{eq:16}\ee   
  where we have made use of eqs. (\ref{eq:9}), (\ref{eq:12}) and (\ref{eq:13}). We see that $h_{\mu\nu}(r)$ transforms as a Wilson line, with $h_{\mu\nu}(x_1)$ and $h_{\mu\nu}(x_2)$  the holonomy factors.  

Let us now define conformal contravariant vectors of weight $\Delta$, transforming under inversion as
    \be
    V^\mu(x)\to \bar V^\mu(\bar x)=\left(\frac 1 {x^2}\right)^{\Delta}h^\mu_\nu(x)V^\nu(x)\label{eq:17}\ee
    and analogous definition for a conformal covariant vector { of weight $\Delta$}
    \be
    W_\mu(x)\to \bar W_\mu(\bar x)=
    {  \left(\frac 1{x^2}\right)^{\Delta}}
    h_\mu^\nu(x)W_\nu(x)\label{eq:18}\ee
    
   Quite clearly $dx^\mu$ and $\de/\de x^\mu$ are contravariant and covariant vectors of weight 1 and -1 respectively.
    Due to the transformation properties of the tensor $h_{\mu\nu}(r)$, it is clear that we can  use it form conformal scalar quantities , since it performs a parallel transport, that is
    \be
    \bar V^\mu(\bar x_1)\bar h_{\mu\nu}(\bar r)\bar V^\nu(\bar x_2)=
    \left(\frac 1{x_1^2 x_2^2}\right)^\Delta V^\mu(x_1)h_{\mu\nu}(r)V^\nu(x_2)\equiv   \left(\frac 1{x_1^2 x_2^2}\right)^\Delta V(x_1)h V(x_2)\label{eq:19}.\ee
   In particular
\be
     {d\bar x}_1^\mu\bar h_{\mu\nu}(\bar r){d\bar x}_2^\nu=\frac{ dx_1^\mu h_{\mu\nu}(r)d x_2^\nu}{x_1^2 x_2^2}\label{eq:20}\ee    
 An analogous result for the covariant vectors of weight $\Delta$ { holds.}

\section{General conformal invariant action}

We start considering   only the space-time variables, $x_1,x_2$ and their first time derivatives, and  let us recall that under inversion:

\be
    \dot{\bar x}_i^2=\frac{\dot x_i^2}{x_i^4},~~~\bar r^2= \frac{r^2}{x_1^2 x_2^2},~~~\dot{\bar x}_1^\mu\bar h_{\mu\nu}\dot{\bar x}_2^\nu=\frac{\dot x_1^\mu h_{\mu\nu}\dot x_2^\nu}{x_1^2 x_2^2}\equiv \frac{\dot x_1 h\dot x_2}{x_1^2x_2^2},~~~i=1,2\label{eq:21}\ee
    
    Since we want to require Diff invariance, we look for the most general expression transforming as a first derivative with respect to the evolution parameter $\tau$, Furthermore we impose  conformal invariance.  By simple dimensional analysis we find 
\be
 L=\left(\frac{\dot x_1^2\dot x_2^2}{r^4}\right)^{1/4} f\left(\frac{\dot x_1 h\dot x_2}{\sqrt{\dot x_1^2\dot x_2^2}}\right)\label{eq:22}\ee
where $f$ is an arbitrary function.  Notice that if we choose the function $f$ as a constant we get the lagrangian discussed in  \cite{Casalbuoni:2014ofa}.

We know from \cite{Casalbuoni:2014ofa} that a free conformal invariant action, describing a single particle depending only on  space-time variables, does not exist. If one wants to describe the theory as a free term plus an interaction, then, one has to introduce the einbeins.
Then, we write the free term in the form
\be
L_{free}= \frac{\dot x_1^2}{2e_1}+ \frac{\dot x_2^2}{2e_2}\label{eq:24}\ee
This allows us to determine the relevant transformation properties of the einbeins. Precisely
\begin{itemize}
\item Under Diff the einbeins transform as time derivatives.
\item Under inversion: $e_i\to e_i/x_i^4$
\end{itemize}
Then, we write down the most general invariant interaction term assuming symmetry under the exchange of the two particles. The result is
\be
L_I=\left(\frac{\sqrt{e_1e_2}}{r^2}\right)F\left(\frac{\dot x_1 h\dot x_2}{\sqrt{\dot x_1^2\dot x_2^2}},\frac{e_1e_2}{\sqrt{\dot x_1^2\dot x_2^2}\,r^2}\right)\label{eq:25}\ee
where $F$ is an arbitrary function of its two arguments which are both conformal and Diff invariant.
On the other hand, if we are interested in finding an explicit relation between the two formulations, we should notice that there is a large arbitrariness in choosing the dependence of the function $F$ from its second argument. For instance, let us consider  the case in which the function $F$ is the product of a power of the second argument times an arbitrary function of the first one, that is
\be
F\left(\frac{\dot x_1 h\dot x_2}{\sqrt{\dot x_1^2\dot x_2^2}},\frac{e_1e_2}{\sqrt{\dot x_1^2\dot x_2^2}\,r^2}\right)=\frac 12\left(\frac{e_1e_2}{\sqrt{\dot x_1^2\dot x_2^2}\,r^2}\right)^{n-1/2}\tilde F\left(\frac{\dot x_1 h\dot x_2}{\sqrt{\dot x_1^2\dot x_2^2}}\right)\label{eq:27}\ee

 The exponent $n-1/2$ has been chosen for simplicity reasons in order to have in the expression of the corresponding $L_I$ the einbeins to the $n$-th power.

The total lagrangian is
 \be
 L=L_{free}+L_I= \frac{\dot x_1^2}{2e_1}+ \frac{\dot x_2^2}{2e_2}+\frac{(e_1e_2)^n}{2r^2}
 \left(\frac 1{\sqrt{\dot x_1^2\dot x_2^2}\,r^2}\right)^{n-1/2}\tilde F\left(\frac{\dot x_1 h\dot x_2}{\sqrt{\dot x_1^2\dot x_2^2}}\right)\label{eq:28}  \ee
 and by varying it, we get two equations for $e_1$ and $e_2$. By solving these equations we find
 \be
 e_1= \left[\left(\frac{\dot x_1^2}{\dot x_2^2}\right)^n\frac {\dot x_1^2}n\,g\right]^{1/(2n+1)},~~~ 
 e_2= \left[\left(\frac{\dot x_2^2}{\dot x_1^2}\right)^n\frac {\dot x_2^2}n\,g\right]^{1/(2n+1)}\label{eq:29}\ee
where we have defined
 \be 
 g= \frac 1{r^2}\left[\frac 1{\sqrt{\dot x_1^2\dot x_2^2}r^2}\right]^{n-1/2} \tilde F\left(\frac{\dot x_1 h\dot x_2}{\sqrt{\dot x_1^2\dot x_2^2}}\right)\label{eq:30}\ee
 Substituting the expressions for the einbeins inside the total lagrangian, we find
 \be
 L=\frac{(1+2n)}2{n}^{-2n/(1+2n)}\left(\frac{\dot x_1^2\dot x_2^2}{r^4} \right)^{1/4}\tilde F\left(\frac{\dot x_1 h\dot x_2}{\sqrt{\dot x_1^2\dot x_2^2}}\right)^{1/(2n+1)}\label{eq:31}\ee
 and, comparing this expression with eq. (\ref{eq:22}) we get
 \be
 \tilde F=\left(\frac 2{2n+1}\right)^{2n+1}n^{2n}f^{2n+1}\label{eq:32}\ee
 This result holds also for $n=0$, noticing that the solutions for the einbeins go to infinity, and
 defining $\lim_{n\to 0} n^{-2n/(1+2n)}=1$.
 
 We see that for any choice of $n$ it is always possible to choose a function $\tilde F$ such to reproduce the lagrangian in eq. (\ref{eq:22}). On the other hand, for an arbitrary choice of $F$ it is practically impossible to find explicitly its relation with $f$. Given this arbitrariness, one can limit himself to the simplest choice $n=1/2$.
 In this case the relation between $f$ and $\tilde F$ is very simple
 \be
 \tilde F=\frac 12 f^2\ee
 
Furtermore, it is useful to consider, two particular cases. The first one corresponds to the model described in 
 \cite{Casalbuoni:2014ofa}, that is to the choice $f =\alpha$
and  correspondingly $\tilde F  =\frac 12 \alpha^2$. Therefore  $F=\frac {\alpha^2}4$,
  in agreement with  \cite{Casalbuoni:2014ofa}.  
  
  We recall that in this case the Diff invariance gives rise to the constraint
  \be
  p_1^2p_2^2=\frac{\alpha^4}{16}\frac 1 {r^4}\label{eq:35}\ee
  
The second  choice is to take $f$ proportional to the square root of its argument, that is
\be
L=\alpha\sqrt{\frac{\dot x_1 h\dot x_2}{r^2}}\label{eq:23}\ee
By choosing $F$ equal to his first argument (always for $n=1/2$), that is
\be
L_I=\frac{\alpha^2}4\left(\frac{\sqrt{e_1e_2}}{r^2}\right)\frac{\dot x_1 h\dot x_2}{\sqrt{\dot x_1^2\dot x_2^2}}\label{eq:26}\ee
and eliminating the einbeins we get back the lagrangian (\ref{eq:23}).
By evaluating the momenta from eq. (\ref{eq:23}) we get
\be
p_1^\mu= \frac 12 \frac {h^{\mu\nu}\dot x_{2\nu}}{\dot x_1 h\dot x_2}l,~~~p_2^\mu= \frac 12 \frac {h^{\mu\nu}\dot x_{1\nu}}{\dot x_1 h\dot x_2}L\ee
From these expression we obtain  the constraint associated to the Diff invariance
\be p_1h p_2= \frac 1 4 \frac{\alpha^2}{r^2}\label{eq:39}\ee

 \section{Study of the constraint}

 In the previous Section we have discussed the constraint from the Diff invariance in two particular cases. We will discuss now the general situation.
Let us start evaluating the momenta from the lagrangian (\ref{eq:22}). They are given by the following expressions
 \bea
 p_1^\mu&=&\frac 12\frac f{\gamma^{3/4}} \frac{\dot x_1^\mu \dot x_2^2}{r^2}+\gamma^{1/4}\frac{h^{\mu\nu}\dot x_{2\nu}-(\dot x_1h\dot x_2) \dot x_1^\mu/\dot x_1^2}{\sqrt{\dot x_1^2\dot x_2^2}}\frac{df}{d\beta}\nn\\
  p_2^\mu&=&\frac 12\frac f{\gamma^{3/4}} \frac{\dot x_2^\mu \dot x_1^2}{r^2}+\gamma^{1/4}\frac{h^{\mu\nu}\dot x_{1\nu}-(\dot x_1h\dot x_2) \dot x_2^\mu/\dot x_2^2}{\sqrt{\dot x_1^2\dot x_2^2}}\frac{df}{d\beta} 
 \eea
 where
 \be
\beta=\frac{\dot x_1 h\dot x_2}{\sqrt{\dot x_1^2\dot x_2^2}},~~~ \gamma= \frac{\dot x_1^2\dot x_2^2}{r^4}\ee
 After some algebra we can find the following expressions
 \bea
p_1hp_2&=&\frac 1{r^2}\left[\frac 1 4\beta f^2+(1-\beta^2)f\frac{df}{d\beta}-\beta(1-\beta^2)\left(\frac{df}{d\beta}\right)^2\right]\nn\\
p_1^2 p_2^2&=&\frac 1 {r^4}\left[\frac 1 4 f^2+(1-\beta^2)\left(\frac{df}{d\beta}\right)^2\right]^2\label{eq:42}\eea 
 Since the terms in the squared brackets depend only on $\beta$, in principle it is possible to eliminate the $\beta$ dependence, for a given $f$, and find a relation involving $p_1hp_2$, $p_1^2p_2^2$ and $r^2$ which is the constraint that we were looking for. In practice, this cannot be done, unless we give explicitly the function $f(\beta)$. 
 Something more can be said by taking $f=\alpha\beta^k$. Then, we get the following ratio
 \be
 \frac{p_1hp_2}{\sqrt{p_1^2p_2^2}}=\pm \frac{(k-1/2)^2\beta^3 +k(1-k)\beta}{(1/4-k^2)\beta^2-k^2}\ee
This could be solved in $\beta$ and then substituted inside one of the eqs. (\ref{eq:42}), obtaining in this way the constraint. It is quite clear the the relation between the momenta will not be polynomial. The only two cases in which we get a polynomial constraint are the ones
 discussed before. In the first case, $f=\alpha$ independent on $\beta$. The second equation decouples from the first one and gives directly  the constraint (\ref{eq:35}).
 In the second particular case where we have taken $f=\alpha\sqrt{\beta}$,   the $\beta$ dependence disappears from the first of the eqs. (\ref{eq:42}) and we get directly the constraint
 (\ref{eq:39}). Also in this case the two equations decouple.

 \section{Conclusions and outlook}
 
 We have constructed the most general action of two interacting relativistic particles invariant under conformal symmetry.
 The construction makes use of a conformal invariant tensor $h_{\mu\nu}$ that transforms as a Wilson line. This fact allow us to connect the velocities of the two particles at different points.
 The lagrangian contains an arbitrary function $f$ (see eq. (\ref{eq:22}).
 We also considered the formulation of the model in terms of enbein variables. 
 
 The mass-shell constraint is general non-polynomial. However there are two particular cases of the model that have polynomial constraints, one is 
 quartic  (see eq. (\ref{eq:35})) and was introduced in \cite{Casalbuoni:2014ofa} and  the other is quadratic (see eq. (\ref{eq:39})).

 It will be interesting to see the possible connection of these constraints with wave functions appearing in reference
\cite{Mintun:2014gua} in the study of higher spin field theories
 
 The extension of the previous results to the case of two superparticles \cite{Casalbuoni:1976tz} invariant under superconformal transformations will be considered elsewhere.

\acknowledgements

We acknowledge 
Jaume Gomis for helpful comments. One of us (J.G.) acknowledges  the hospitality at the Dipartamento di Fisica de la Universita di Firenze where this work has been finished. Partially financed by  INFN
and  FPA2013-46570,  2014-SGR-104, CPAN, Consolider CSD 2007-0042.

\end{document}